\documentstyle[12pt]{article}
\begin{document}
\begin{titlepage}
\begin{flushright}
ICTP--94--35
\end{flushright} \quad\\
\vspace{1.8cm}
\begin{center}
{\bf \LARGE Bethe Ansatz and Quantum Davey-Stewartson 1 System with
Multicomponent in Two Dimensions}
\footnote{\footnotesize
The work was supported in part by National
Science Foundation and Science Academy
Foundation in P.R.China.}
\\
\vspace{1cm}
{\bf Yi Cheng}
\\
%\bigskip
Department of Mathematics\\
%$^{2}$Fundamental Physics Center \\
University of Science and Technology of China\\
Hefei Anhui 230026, P.R.China
\\
\vspace{.5cm}
{\bf Mu-Lin Yan}\footnote{\footnotesize On leave from
Fundamental Physics Center,
University of Science and Technology of China,
Hefei, Anhui 230026, P.R.China}
\\
International Center for Theoretical Physics\\
P.O.Box 586, 34100 Trieste, Italy
\\
\vspace{.5cm}
{\bf Bao-Heng Zhao} \\
Department of Physics\\
 Graduate School, Chinese Academy of Sciences \\
 P.O.Box 3908, Beijing 100039, P.R.China\\
\vspace{1cm}
%\date{Feb. 13, 1994}
Feb. 11, 1994 \\
 \vspace{1cm} {\bf Abstract:} \\
 \parbox[t]{\textwidth}{
The quantum 2-component DS1 system was reduced to two 1D many-body
problems with $\delta-$function interactions, which were solved by
Bethe ansatz. Using the ansatz in ref.[1] and introducing symmetric
and antisymmetric Young operators of the permutation group, we obtain
the exact solutions for the system.
} \end{center}\end{titlepage} \newpage

%\section{Introduction}
  1, The Davey-Stewartson 1 (DS1) system is an integrable model in
space of two spatial and one temporal dimensions ((2+1)D). The
quantized DS1 system can be formulated in terms of Hamiltonians
of quantum
many-body problems in two dimensions, and some of them can be solved
exactly$^{[1][2]}$.
Particularly, it has been shown in ref.[2]
that these 2D quantum many-body problems can
be reduced to the solvable one-dimensional quantum many-body problems
with two-body potentials$^{[3]}$. Thus through solving the 1D many-body
problems we can get the solutions of 2D's. In the present paper, we
intend to generalize this idea to multicomponent DS1 system.
Specifically, we will consider the case that the potential between two
particles with two components in one dimension is delta-function. It is
well known that the Bethe ansatz is useful and legitimate for solving
1D many-body problems with delta-function interactions$^{[4][5]}$. Thus
Bethe ansatz (including nested Bethe ansatz or Bethe-Yang
ansatz$^{[5]}$) will also play important role for solving the 2D
many-body problems induced from multicomponent DS1 system. Like 1D
cases, the effects of
multicomponent in 2D many-body problems can not be trivially counted by
suming single-component modes. Namely, a careful and non-trivial
consideration is necessary. For definiteness, we will do so for
a specific model of 2D quantum DS1 system with two components.

   2, Following usual DS1 equation$^{[1][6]}$, the equation for the DS1
system with two components reads
\begin{equation}
i{\bf \dot{q}} =-\frac{1}{2} (\partial _{x}^{2}
+\partial_{y}^{2}){\bf q}+
 iA_{1}{\bf q}+iA_{2}{\bf q},
\end{equation}
where ${\bf q}$ has two colour components,
\begin{equation}
{\bf q}=
    \left( \begin{array}{c}
    q_{1}\\
    q_{2}\\
    \end{array} \right),
\end{equation}
and
\begin{eqnarray*}
(\partial _{x}-\partial
_{y})A_{1}&=&-ic(\partial_{x}+\partial_{y})({\bf q^{\dag}q})\\
(\partial _{x}+\partial
_{y})A_{2}&=&ic(\partial_{x}-\partial_{y})({\bf q^{\dag}q})
\end{eqnarray*}
where notation ${\bf \dag }$ means the hermitian transposition,
and $c$ is the coupling constant.
Introducing the coordinates $\xi =x+y, \eta =x-y,$ we have
\begin{eqnarray}
A_{1}&=&-ic\partial _{\xi}\partial _{\eta}^{-1}({\bf q^{\dag}q})
        -iu_{1}(\xi)  \\
A_{2}&=&ic\partial _{\eta}\partial _{\xi}^{-1}({\bf q^{\dag}q})
        +iu_{2}(\eta)
\end{eqnarray}
where
\begin{equation}
\partial_{\eta}^{-1}({\bf
q^{\dag}q})=\frac{1}{2}(\int_{-\infty}^{\eta} d\eta^{\prime}-
\int^{\infty}_{\eta} d\eta^{\prime}){\bf q^{\dag}}(\xi,\eta^{\prime},t)
{\bf q}(\xi,\eta^{\prime},t),
\end{equation}
and $u_{1}$ and $u_{2}$ are constants of integration. According to ref.[2],
we choose them as
\begin{eqnarray}
u_{1}(\xi)&=&\frac{1}{2}\int d\xi^{\prime}d\eta^{\prime}U_{1}(\xi-\xi^{\prime})
           {\bf q^{\dag}}(\xi^{\prime},\eta^{\prime},t)
               {\bf q}(\xi^{\prime},\eta^{\prime},t)  \\
u_{2}(\eta)&=&\frac{1}{2}\int
d\xi^{\prime}d\eta^{\prime}U_{2}(\eta-\eta^{\prime})
           {\bf q^{\dag}}(\xi^{\prime},\eta^{\prime},t)
               {\bf q}(\xi^{\prime},\eta^{\prime},t).
\end{eqnarray}
Thus eq.(1) can be written as
\begin{eqnarray}
i{\bf \dot{q}}=-(\partial_{\xi}^{2}+\partial_{\eta}^{2})
  {\bf q}+c[\partial_{\xi}\partial_{\eta}^{-1}({\bf q^{\dag}q})
+\partial_{\eta}\partial_{\xi}^{-1}({\bf q^{\dag}q})]{\bf q}
\nonumber \\
+\frac{1}{2}\int d\xi^{\prime}d\eta^{\prime}[U_{1}(\xi-\xi^{\prime})+
U_{2}(\eta-\eta^{\prime})]({\bf q}^{\dag \prime}{\bf
q}^{\prime}){\bf q},
\end{eqnarray}
where ${\bf q}^{\prime}={\bf q}(\xi^{\prime},\eta^{\prime},t)$. We
quantize the system with the canonical commutation relations
%\begin{eqnarray}
\begin{equation}
 [ q_{a}(\xi,\eta,t), q^{\dag}_{b}(\xi^{\prime},
\eta^{\prime},t)]_{\pm}
= 2\delta_{_{ab}}\delta(\xi-\xi^{\prime})\delta(\eta-\eta^{\prime}), \\
\end{equation}
\begin{equation}
 [ q_{a}(\xi,\eta,t), q_{b}(\xi^{\prime},\eta^{\prime},t)]_{\pm}
=0.
\end{equation}
%\end{eqnarray}
where $a,b = 1$ or $2$, $[ , ]_{+}$ and $[ , ]_{-}$ are anticommutator
 and commutator respectively.
Then eq.(8) can be written in the form
\begin{equation}
\dot{{\bf q}}=i[H,{\bf q}]
\end{equation}
where $H$ is the Hamiltonian of the system
\begin{eqnarray}
H=\frac{1}{2}\int d\xi d\eta \begin{array}{c}${\LARGE(}$\end{array}
 -{\bf q}^{\dag}
(\partial_{\xi}^{2}+\partial_{\eta}^{2})
  {\bf q}+\frac{c}{2}{\bf q}^{\dag}[(\partial_{\xi}\partial_{\eta}^{-1}
  +\partial_{\eta}\partial_{\xi}^{-1})({\bf q^{\dag}q})]{\bf q}
\nonumber \\
+\frac{1}{4}\int d\xi^{\prime}d\eta^{\prime}{\bf q}^{\dag}
[U_{1}(\xi-\xi^{\prime})
+U_{2}(\eta-\eta^{\prime})]({\bf q}^{\prime \dag}{\bf
q}^{\prime}){\bf q}\begin{array}{c}${\LARGE)}$\end{array}.
\end{eqnarray}
The $N$-particle eigenvalue problem is
\begin{equation}
H\mid \Psi \rangle =E\mid \Psi \rangle
\end{equation}
where
%\begin{eqnarray}
\begin{equation}
\mid \Psi \rangle=\int d\xi_{1}d\eta_{1}\ldots d\xi_{N}d\eta_{N}
\sum_{a_{1}\ldots a_{N}} \Psi_{a_{1}\ldots a_{N}}(\xi_{1}\eta_{1}
\ldots \xi_{N}\eta_{N})
% \nonumber  \\
{\bf q}^{\dag}_{a_{1}}(\xi_{1}\eta_{1})\ldots
{\bf q}^{\dag}_{a_{N}}(\xi_{N}\eta_{N}) \mid 0 \rangle.
%\end{eqnarray}
\end{equation}
The $N$-particle wave function $ \Psi_{a_{1}\ldots a_{N}}$ is defined
by eq.(14), which satisfies the $N$-body Schrodinger equation
\begin{eqnarray}
-\sum_{i}(\partial_{\xi_{i}}^{2}+\partial_{\eta_{i}}^{2})
 \Psi_{a_{1}\ldots a_{N}}+c\sum_{i<j}
 [\epsilon (\xi_{ij})\delta^{\prime}(\eta_{ij})
  +\epsilon (\eta_{ij})\delta^{\prime}(\xi_{ij})] \Psi_{a_{1}\ldots a_{N}}
  \nonumber  \\
  +\sum_{i<j}[U_{1}(\xi_{ij})+U_{2}(\eta_{ij})] \Psi_{a_{1}\ldots a_{N}}
  =E \Psi_{a_{1}\ldots a_{N}}
\end{eqnarray}
where $\xi_{ij}=\xi_{i}-\xi_{j}, \delta^{\prime}(\xi_{ij})=\partial_{\xi_{i}}
\delta (\xi_{ij})$, and $\epsilon (\xi_{ij})=1$ for $\xi_{ij}>0, 0$ for
$\xi_{ij}=0, -1$ for $\xi_{ij}<0.$ Since there are products of
distributions in eq.(15), an appropriate regularezation for avoiding
uncertainty is necessary. This issue has been discussed in ref.[7].

  3, Our purpose is to solve the $N$-body Schrodinger equation (15). The
results in ref.[2] remind us that we can make the following ansatz
\newpage
\begin{eqnarray}
\Psi_{a_{1}\ldots a_{N}}
&=&\sum_{_{\begin{array}{ccc}
       a_{1}^{\prime} & \ldots & a_{N}^{\prime} \\
       b_{1}^{\prime} & \ldots & b_{N}^{\prime} \end{array}}}
 \prod_{i<j} (1-\frac{c}{4}\epsilon (\xi_{ij})\epsilon (\eta_{ij}))
 \nonumber  \\
& &\times {\cal M}_{a_{1}\ldots a_{N}, a_{1}^{\prime}\ldots a_{N}^{\prime}}
 {\cal N}_{a_{1}\ldots a_{N}, b_{1}^{\prime}\ldots b_{N}^{\prime}}
X_{ a_{1}^{\prime}\ldots a_{N}^{\prime}}(\xi_{1}\ldots \xi_{N})
Y_{ b_{1}^{\prime}\ldots b_{N}^{\prime}}(\eta_{1}\ldots \eta_{N})
\end{eqnarray}
where ${\cal M}$ and ${\cal N}$ are matrices independent of the
coordenates of $\xi$ and $\eta$;
$X_{ a_{1}\ldots a_{N}}(\xi_{1}\ldots \xi_{N})$ and
$Y_{ b_{1}\ldots b_{N}}(\eta_{1}\ldots \eta_{N})$ are required to satisfy
the following one-dimensional $N$-body Schrodinger equations with
two-body potentials
\begin{eqnarray}
-\sum_{i} \partial_{\xi_{i}}^{2} X_{a_{1}\ldots a_{N}}
+\sum_{i<j} U_{1}(\xi_{ij}) X_{a_{1}\ldots a_{N}}&=&E_{1} X_{a_{1}\ldots a_{N}}
\\
-\sum_{i} \partial_{\eta_{i}}^{2} Y_{b_{1}\ldots b_{N}}
+\sum_{i<j} U_{2}(\eta_{ij}) Y_{b_{1}\ldots b_{N}}&=&E_{2}Y_{b_{1}\ldots b_{N}}
\end{eqnarray}
where $E_{1}+E_{2}=E$. At this stage ${\cal M}$ and ${\cal N}$ are
unknown temporarily.
 It is expected that after 1D many-body problems (i.e.,
eqs.(17) (18)) are solved, we could construct the solutions $\Psi
_{A_{1} \ldots A_{N}}$ for 2D many-body problems eq.(15) through
constructing an appropriate ${\cal M} \times {\cal N}$-matrix.
% where $\otimes$ means the tenser product.

  Now, let us consider the case of $U_{1}(\xi_{ij})=2g\delta (\xi_{ij})$
and $U_{2}(\eta_{ij})=2g\delta (\eta_{ij})$ ($g>0$, the coupling
constant). Then eqs.(17) and (18) become
\begin{eqnarray}
-\sum_{i} \partial_{\xi_{i}}^{2} X_{a_{1}\ldots a_{N}}
+2g \sum_{i<j} \delta(\xi_{ij}) X_{a_{1}\ldots a_{N}}
&=&E_{1} X_{a_{1}\ldots a_{N}}
\\
-\sum_{i} \partial_{\eta_{i}}^{2} Y_{b_{1}\ldots b_{N}}
+2g \sum_{i<j} \delta(\eta_{ij}) Y_{b_{1}\ldots b_{N}}
&=&E_{2}Y_{b_{1}\ldots b_{N}}.
\end{eqnarray}
If $X$ and $Y$ are wave functions of Fermions with two components,
denoted by $X^{F}$ and $Y^{F}$, the
problem has been solved by Yang $^{[6]}$ (more explicitly, see
ref.[8] and ref.[9]). According to the Bethe ansatz, the continual solution of
eq.(15) in the region of $0<\xi_{Q_{1}}<\xi_{Q_{2}}<\ldots
<\xi_{Q_{N}}<L$ reads
\begin{eqnarray}
X^{F}&=&\sum_{P}
\alpha_{P}^{(Q)}\exp\{i[k_{P_{1}}\xi_{Q_{1}}+\ldots+k_{P_{N}}\xi_{Q_{N}}
]\} \nonumber  \\
&=&\alpha_{_{12\ldots N}}^{(Q)}e^{i(k_{1}\xi_{Q_{1}}+k_{2}\xi_{Q_{2}}
+\ldots+k_{N}\xi_{Q_{N}})}
+\alpha_{_{21\ldots N}}^{(Q)}e^{i(k_{2}\xi_{Q_{1}}+k_{1}\xi_{Q_{2}}
+\ldots+k_{N}\xi_{Q_{N}})}
\nonumber  \\
& &+(N!-2) \begin{array}{c} $others terms$ \end{array}
\end{eqnarray}
where  $X^{F}\in \{ X^{F}_{a_{1}\ldots a_{N}} \}$,
$P=[P_{1},P_{2},\cdots,P_{N}]$
and $Q=[Q_{1},Q_{2},\cdots,Q_{N}]$
are two permutations of the
integers $1,2,\ldots ,N$, and
\begin{eqnarray}
\alpha^{(Q)}_{\ldots ij \ldots}&=&Y^{lm}_{ji} \alpha^{(Q)}_{\ldots ji \ldots},
  \\
Y^{lm}_{ji}&=&\frac{-i(k_{j}-k_{i})P^{lm}+g}{i(k_{j}-k_{i})-g}
\end{eqnarray}
The eigenvalue is given by
\begin{equation}
E_{1}=k_{1}^{2}+k_{2}^{2}+\ldots +k_{N}^{2},
\end{equation}
where $\{ k_{i} \}$ are determined by the Bethe ansatz equations,
\begin{eqnarray}
& &e^{ik_{j}L}=\prod_{\beta =1}^{M} \frac{i(k_{j}-\Lambda_{\beta})-g/2}
{i(k_{j}-\Lambda_{\beta})+g/2}   \\
& &\prod_{j=1}^{N} \frac{i(k_{j}-\Lambda_{\alpha})-g/2}
{i(k_{j}-\Lambda_{\alpha})+g/2}
=-\prod_{\beta =1}^{M} \frac{i(\Lambda_{\alpha}-\Lambda_{\beta})+g}
{i(\Lambda_{\alpha}-\Lambda_{\beta})-g}
\end{eqnarray}
with $\alpha =1,\ldots, M, j=1,\ldots, N.$ Through exactly same
procedures we can get the solution $Y^{F}$ and $E_{2}$ to eq.(20).

    In the Bonsonic case, the wave-functions,
denoted by $X^{B}$ and $Y^{B}$, are given by
\begin{eqnarray}
X^{B}&=&\sum_{P}
\beta_{P}^{(Q)}\exp\{i[k_{P_{1}}\xi_{Q_{1}}+\ldots+k_{P_{N}}\xi_{Q_{N}}
]\}   \\
\beta^{(Q)}_{\ldots ij \ldots}&=&Z^{lm}_{ji} \beta^{(Q)}_{\ldots ji \ldots},
  \\
Z^{lm}_{ji}&=&\frac{i(k_{j}-k_{i})P^{lm}+g}{i(k_{j}-k_{i})-g}
\end{eqnarray}
and the Bethe ansatz equations are as following$^{[9]}$
\begin{eqnarray}
& &e^{ik_{j}L}=-\prod_{i=1}^{N} \frac{k_{j}-k_{i}+ig}{k_{j}-k_{i}-ig}
\prod_{\beta =1}^{M} \frac{\Lambda_{\beta}-k_{j}+g/2}
{\Lambda_{\beta}-k_{j}-g/2}   \\
& &\prod_{\alpha =1}^{M} \frac{\Lambda_{\beta}-\Lambda_{\alpha}+ig}
{\Lambda_{\beta}-\Lambda_{\alpha}-ig}
=-\prod_{j =1}^{N} \frac{\Lambda_{\beta}-k_{j}+ig/2}
{\Lambda_{\beta}-k_{j}-ig/2}.
\end{eqnarray}
Similarly for
$Y^{B}$. It is well known that $X^{F}$ and
$Y^{F}$($X^{B}$ and $Y^{B}$) are antisymmetric (symmetrec) when
the coordinates and the colour-indeces of the
particles are interchanged simultaneously.

    4, For permutation group $S_{N}: \{ e_{i}, i=1,\cdots,
N!\}$, the totally symmetric Young operator is
\begin{equation}
{\cal O}_{N}=\sum_{i=1}^{N!} e_{i},
\end{equation}
and the totally antisymmetric Young operator is
\begin{equation}
{\cal A}_{N}=\sum_{i=1}^{N!} (-1)^{P_{i}}e_{i}.
\end{equation}
The Young diagram for ${\cal O}_{N}$ is
\begin{tabular}{|l|l|l|l|l|r|}  \hline
 1  & 2 &3 &$\cdots$ & N  \\  \hline
\end{tabular},
and for ${\cal A}_{N}$, it is
\begin{tabular}{|l|r|}  \hline
 1  \\  \hline
 2  \\  \hline
 \vdots  \\  \hline
 N  \\   \hline
\end{tabular}.
To $S_{3}$, for example, we have
\begin{equation}
{\cal O}_{3}=1+P^{12}+P^{13}+P^{23}+P^{12}P^{23}+P^{23}P^{12}.
\end{equation}
\begin{equation}
{\cal A}_{3}=1-P^{12}-P^{13}-P^{23}+P^{12}P^{23}+P^{23}P^{12}.
\end{equation}
Lemma 1: $({\cal O}_{N}X_{F})(\xi_{1},\xi_{2},\cdots,\xi_{N})$ is
antisymmetric with respect to the coordinate's interchanges of
$(\xi_{i} \longleftrightarrow \xi_{j})$.

\noindent Proof: From the definition of ${\cal O}_{N}$ (eq.(32)), we
have
\begin{equation}
{\cal O}_{N}P^{ab}=P^{ab}{\cal O}_{N}={\cal O}_{N}.
\end{equation}
To $N=3$ case, for example, the direct calculations show ${\cal
O}_{3}P^{12}=P^{12}{\cal O}_{3}={\cal O}_{3},
{\cal O}_{3}P^{23}=P^{23}{\cal O}_{3}={\cal O}_{3}$
and so on. Using eqs.(36) and (23), we have
\begin{equation}
{\cal O}_{N}Y^{lm}_{ij}=(-1){\cal O}_{N}.
\end{equation}
{}From eqs.(21) and (23), $X^{F}$ can be written as
\begin{eqnarray}
X^{F}&=&\{ e^{i(k_{1}\xi_{Q_{1}}+k_{2}\xi_{Q_{2}}
+\ldots+k_{N}\xi_{Q_{N}})}
+Y^{12}_{12}e^{i(k_{2}\xi_{Q_{1}}+k_{1}\xi_{Q_{2}}
+\ldots+k_{N}\xi_{Q_{N}})}
\nonumber  \\
& &+Y^{23}_{13}Y^{12}_{12}e^{i(k_{2}\xi_{Q_{1}}
+k_{3}\xi_{Q_{2}}+k_{1}\xi_{Q_{3}}
+\ldots+k_{N}\xi_{Q_{N}})}
+(N!-3) \begin{array}{c} $other terms$ \end{array}
\} \alpha_{_{12\ldots N}}^{(Q)}
\end{eqnarray}
Using eqs.(37) and (38), we obtain
\begin{eqnarray}
({\cal O}_{N}X^{F})(\xi_{1},\cdots,\xi_{N})=
\{ e^{i(k_{1}\xi_{Q_{1}}+k_{2}\xi_{Q_{2}}
+\ldots+k_{N}\xi_{Q_{N}})}
-e^{i(k_{2}\xi_{Q_{1}}+k_{1}\xi_{Q_{2}}
+\ldots+k_{N}\xi_{Q_{N}})}
\nonumber   \\
+e^{i(k_{2}\xi_{Q_{1}}
+k_{3}\xi_{Q_{2}}+k_{1}\xi_{Q_{3}}
+\ldots+k_{N}\xi_{Q_{N}})}
+(N!-3)\begin{array}{c}$other terms$\end{array}
\} {\cal O}_{N} \alpha_{_{12\ldots N}}^{(Q)}
\nonumber   \\
=\sum_{P} (-1)^{P}
\exp\{i[k_{P_{1}}\xi_{Q_{1}}+\ldots+k_{P_{N}}\xi_{Q_{N}}
]\}({\cal O}_{N} \alpha_{_{12\ldots N}}^{(Q)}).
\end{eqnarray}
Therefore we conclude that $({\cal O}_{N}X^{F})(\xi_{1},\cdots,\xi_{N})$
is antisymmetric
with respect to $(\xi_{i}\longleftrightarrow \xi_{j})$.

\noindent Lemma 2: $({\cal A}_{N}X^{B})(\xi_{1},\xi_{2},\cdots,\xi_{N})$ is
antisymmetric with respect to the coordinate's interchanges of
$(\xi_{i} \longleftrightarrow \xi_{j})$.

\noindent  Proof: Noting (see eqs.(33) (29) (27))
\begin{eqnarray}
{\cal A}_{N} P^{ab}=P^{ab}{\cal A}=-{\cal A}_{N},  \\
{\cal A}_{N} Z^{lm}_{ij}=(-1){\cal A}_{N},
\end{eqnarray}
we then have
\begin{equation}
({\cal A}_{N}X^{B})(\xi_{1},\cdots,\xi_{N})
=\sum_{P} (-1)^{P}
\exp\{i[k_{P_{1}}\xi_{Q_{1}}+\ldots+k_{P_{N}}\xi_{Q_{N}}
]\}({\cal A}_{N} \beta_{_{12\ldots N}}^{(Q)}).
\end{equation}
Then the Lemma is proved.

    5, The ansatz of eq.(16) can be compactly written as
\begin{equation}
\Psi
= \prod_{i<j} (1-\frac{c}{4}\epsilon (\xi_{ij})\epsilon (\eta_{ij}))
  ({\cal M}X)  ({\cal N}Y)
\end{equation}
where $({\cal M}X)$ and $({\cal N}Y)$ are required to be
antisymmetric under the interchanges of the coordinate vairables.
According to Lemmas 1 and 2, we see that
\begin{equation}
 {\cal M},{\cal N} =\left\{
\begin{array}{ccc}{\cal O}_{N} & &$for 1D Fermion$  \\
                  {\cal A}_{N} & &$for 1D Boson.$
\end{array} \right.
\end{equation}
If the DS1 fields $q_{a}(\xi , \eta)$ in eq.(1) are (2+1)D Bose fields,
the commutators ($[ , ]_{-}$, see (9) and (10))
must be used to quantized
the system and the 2D many-body wave functions denoted by $\Psi ^{B}$
must be symmetric under the simultaneous
colour-interchang $(a_{i} \longleftrightarrow
a_{j})$ and the coordinate-interchange $((\xi_{i} \eta_{i})
\longleftrightarrow  (\xi_{j} \eta_{i}))$. Namely, the 2D Bose wave
functions $\Psi^{B}$ must satisfy
\begin{equation}
P^{a_{i} a_{j}}\Psi^{B}\mid _{\xi_{i}\eta_{i} \longleftrightarrow
\xi_{j} \eta_{j}} =\Psi^{B}.
\end{equation}
For $q_{a}$ ,(2+1)D Fermi fields, the anticommutators should be used,
and $\Psi^{F}$ must be antisymmetric under
simultaneous interchange of
$(a_{i} \longleftrightarrow
a_{j})$ and $((\xi_{i} \eta_{i})
\longleftrightarrow  (\xi_{j} \eta_{i}))$, namely,
\begin{equation}
P^{a_{i} a_{j}}\Psi^{F}\mid _{\xi_{i}\eta_{i} \longleftrightarrow
\xi_{j} \eta_{j}} =-\Psi^{F}.
\end{equation}
Thus for the 2D Boson case, two solutions of $\Psi^{B}$ can be
constructed as following
\begin{eqnarray}
\Psi^{B}_{1}
= \prod_{i<j} (1-\frac{c}{4}\epsilon (\xi_{ij})\epsilon (\eta_{ij}))
 [ {\cal O_{N}}X^{F}(\xi_{1}\cdots \xi_{N})]
 [ {\cal O_{N}}Y^{F}(\eta_{1}\cdots \eta_{N})], \\
\Psi^{B}_{2}
= \prod_{i<j} (1-\frac{c}{4}\epsilon (\xi_{ij})\epsilon (\eta_{ij}))
 [ {\cal A_{N}}X^{B}(\xi_{1}\cdots \xi_{N})]
  [{\cal A_{N}}Y^{B}(\eta_{1}\cdots \eta_{N})].
\end{eqnarray}
Using eqs.(36),(39),(40) and (42), we can check eq.(45) diractly. In
addition, from the Bethe ansatz equations (25) (26) (30) (31) and $E=
E_{1}+E_{2}$, we can see that the eigenvalues of $\Psi^{B}_{1}$ and
$\Psi^{B}_{2}$ are different from
each other generally, i.e., the states
corrosponding to $\Psi^{B}_{1}$ and $\Psi^{B}_{2}$ are non-degenerate.

For the 2D Fermion case, the desired results are
\begin{eqnarray}
\Psi^{F}_{1}
= \prod_{i<j} (1-\frac{c}{4}\epsilon (\xi_{ij})\epsilon (\eta_{ij}))
 [ {\cal O_{N}}X^{F}(\xi_{1}\cdots \xi_{N})]
 [ {\cal A_{N}}Y^{B}(\eta_{1}\cdots \eta_{N})], \\
\Psi^{F}_{2}
= \prod_{i<j} (1-\frac{c}{4}\epsilon (\xi_{ij})\epsilon (\eta_{ij}))
 [ {\cal A_{N}}X^{B}(\xi_{1}\cdots \xi_{N})]
  [{\cal O_{N}}Y^{F}(\eta_{1}\cdots \eta_{N})].
\end{eqnarray}
Eq.(46) can also be checked diractly. The eigenvalues corresponding to
$\Psi^{F}$ are also determined by the Bethe equations and
$E=E_{1}+E_{2}$.

 The proof given in ref.[2] can be extended to show that
$\Psi^{B}_{1,2}$ and
$\Psi^{F}_{1,2}$ , shown in above, are also
the exact solutions of eq.(15). Thus
 we conclude that the 2D quantum
many-body problem induced from the quantum DS1 system with
2-component has been solved exactly.

6, To summarize. We formulated the quantum multicomponent DS1 system in
terms of the quantum multicomponent many-body Hamiltonain in 2D space.
Then we reduced this 2D Hamiltonain to two 1D multicomponent many-body
problems. As the potential between two particles with two components in
one dimension is $\delta-$function, the Bethe ansatz was used to solve
these 1D problems. By using the ansatz of ref.[1] and introducing some
useful Young
operators, we presented a new ansatz for fusing two 1D-solutions
to construct 2D wave functions of the quantum many-body problem which is
induced from the quantum 2-component DS1 system. There are two types of
wave functions: Boson's and Fermion's. Both of them satisfy the 2D many-body
Schrodinger equation of the DS1 system exactly.

{\bf Acknowledgment:} One of us (MLY) is grateful to
S. Randjbar-Daemi for discussions.

%\end{document}

\vskip2.5cm

%\end{document}

%\newpage

%\newpage
%{\Large\bf Caption}
%\vskip1cm
%\begin{description}
%\item[Fig.1] \ The curve for the function of $m$ to $e$ (eq.(38)).  Both $m$
%and $e$ are the parameters in the model (eq.(8)).
%\end{description}
\end{document}